# A variant of the Brillouin-Wigner perturbation theory with Epstein-Nesbet partitioning

Sangyoub Lee,[1,a)] Cheol Ho Choi,[2] Eunji Kim,[1] and Young Kyun Choi[1]

[1]*Department of Chemistry, Seoul National University, Seoul 151-747, South Korea*

[2]*Department of Chemistry, Kyungpook National University, Taegu 702-701, South Korea*

We present an elementary pedagogical derivation of the Brillouin-Wigner and the Rayleigh-Schrödinger perturbation theories with Epstein-Nesbet partitioning. A variant of the Brillouin-Wigner perturbation theory is also introduced, which can be easily extended to the quasi-degenerate case. A main advantage of the new theory is that the computing time required for obtaining the successive higher-order results is negligible after the third-order calculation. We illustrate the accuracy of the new perturbation theory for some simple model systems like the perturbed harmonic oscillator and the particle in a box.

---

[a)] Corresponding author. E-mail: sangyoub@snu.ac.kr

## I. INTRODUCTION

One of the most useful approximation methods in quantum mechanics is the perturbation theory.[1] There are many variations of the perturbation theory for dealing with stationary bound systems. Of these, the Rayleigh-Schrödinger perturbation theory (RSPT) and the Brillouin-Wigner perturbation theory (BWPT) are the most fundamental and general ones.[2,3] Some theories may give more accurate results but their applicability is limited to specific systems. In this paper, we present an elementary derivation of BWPT and RSPT with Epstein-Nesbet (EN) partitioning,[4,5] and introduce a new variant of the BWPT. A major advantage of the new perturbation theory is that the computing time required for obtaining the successive higher-order results beyond the second-order one increases linearly. For simple model systems like the perturbed harmonic oscillator and the particle in a box, it is shown that the ground-state energies calculated from the new perturbation theory have comparable accuracy as those calculated up to the same order by BWPT and RSPT with EN partitioning.

## II. NONDEGENERATE PERTURBATION THEORY

Let us consider a system with the Hamiltonian given by $H = H_0 + \lambda V$. The unperturbed energies and the state vectors are denoted by $E_m^{(0)}$ and $\left| \psi_m^{(0)} \right\rangle$, respectively. We assume that the zeroth-order states constitute a complete and orthonormal basis. Then the perturbed state vector of the $n$th state of interest can be represented as

$$\left| \psi_n \right\rangle = \sum_m c_{mn} \left| \psi_m^{(0)} \right\rangle = c_{nn} \left( \left| \psi_n^{(0)} \right\rangle + \sum_{m \neq n} c'_{mn} \left| \psi_m^{(0)} \right\rangle \right), \tag{1}$$

where $c_{mn} = \left\langle \psi_m^{(0)} \middle| \psi_n \right\rangle$ and $c'_{mn} \equiv c_{mn} / c_{nn}$, and the Schrödinger equation for the perturbed Hamiltonian can be represented as a matrix equation,

$$\sum_m (H_{lm} - E_n \delta_{lm}) c'_{mn} = 0 \quad \text{for all } l. \tag{2}$$

The Hamiltonian matrix element can be written as

$$H_{lm} = \left\langle \psi_l^{(0)} \middle| H_0 + \lambda V \middle| \psi_m^{(0)} \right\rangle = E_m^{(0)} \delta_{lm} + \lambda V_{lm} \tag{3}$$

with $V_{lm} = \left\langle \psi_l^{(0)} \middle| V \middle| \psi_m^{(0)} \right\rangle$. In the EN partitioning, the full Hamiltonian matrix is partitioned as

$$H_{lm} = E'_m \delta_{lm} + \lambda V'_{lm} \quad \text{with} \quad E'_m = E_m^{(0)} + \lambda V_{mm} \quad \text{and} \quad V'_{lm} = V_{lm}(1 - \delta_{lm}). \tag{4}$$

By substituting Eq. (4) into Eq. (2), we obtain

$$(E'_l - E_n) c'_{ln} + \sum_{m \neq l} \lambda V'_{lm} c'_{mn} = 0 \quad \text{for all } l \tag{5}$$

Consider first the simpler case in which the energy of interest, $E_n$, is well separated from $E'_l$ for all $l$ other than $n$. For $l = n$, Eq. (5) gives

$$E_n = E'_n + \lambda \sum_{m \neq n} V'_{nm} c'_{mn} \tag{6}$$

On the other hand, for $l \neq n$ Eq. (5) gives

$$c'_{mn} = \frac{\lambda}{E_n - E'_m} \left( V'_{mn} + \sum_{m_1 \neq n} V'_{mm_1} c'_{m_1 n} \right) \tag{7}$$

Note that Eqs. (6) and (7) are exact relations. In the RSPT, it is assumed that the energy and the state vector of the perturbed system are analytic functions, so that they can be represented as MacLaurin series in $\lambda$. In Eq. (7), no such assumption is made, but it involves the unknown quantity $E_n$.

By iteration, Eq. (7) gives the following perturbation series expression for $c'_{mn}$ in the BWPT with EN partitioning:

$$c'_{mn} = \lambda \frac{V'_{mn}}{E_n - E'_m} + \lambda^2 \sum_{m_1 \neq n} \frac{V'_{mm_1} V'_{m_1 n}}{(E_n - E'_m)(E_n - E'_{m_1})}$$

$$+ \lambda^3 \sum_{m_1 \neq n} \sum_{m_2 \neq n} \frac{V'_{mm_1} V'_{m_1 m_2} V'_{m_2 n}}{(E_n - E'_m)(E_n - E'_{m_1})(E_n - E'_{m_2})} + \cdots \tag{8}$$

By substituting this expression for $c'_{mn}$ into Eq. (6), we obtain the perturbation series expression for energy in BWPT with EN partitioning:

$$E_n = E'_n + \lambda^2 \sum_{m \neq n} \frac{V'_{nm} V'_{mn}}{E_n - E'_m} + \lambda^3 \sum_{m \neq n} \sum_{m_1 \neq n} \frac{V'_{nm} V'_{mm_1} V'_{m_1 n}}{(E_n - E'_m)(E_n - E'_{m_1})}$$

$$+ \lambda^4 \sum_{m \neq n} \sum_{m_1 \neq n} \sum_{m_2 \neq n} \frac{V'_{nm} V'_{mm_1} V'_{m_1 m_2} V'_{m_2 n}}{(E_n - E'_m)(E_n - E'_{m_1})(E_n - E'_{m_2})} + \cdots \tag{9}$$

The $k$th-order energy of the BWPT is given by truncating the series in Eq. (9) after the $\lambda^k$ term. The resulting equation is solved for $E_n$ to give the $k$th-order energy $E_n^{(BWk)}$; in this paper $E_n^{(BWk)}$ denotes an approximate energy that is correct up to $O(\lambda^k)$, not the $k$th-order correction. A more approximate procedure to evaluate $E_n^{(BWk)}$ is just to put the value $E_n^{(BW,k-1)}$ for $E_n$ in the truncated series on the right hand side.

The RSPT with EN partitioning can be easily derived from Eqs. (6) and (8). The first-order approximation to $c'_{mn}$ in BWPT can be written as

$$c'^{(BW1)}_{mn} = \lambda \frac{V'_{mn}}{E_n - E'_m} = \lambda \frac{V'_{mn}}{E'_n + O(\lambda^2) - E'_m} \cong \lambda \frac{V'_{mn}}{E'_n - E'_m} + O(\lambda^3) \tag{10}$$

From Eqs. (6) and (10), we then obtain the first-order approximation to $c'_{mn}$ and the second-order approximation to $E_n$ in RSPT with EN partitioning as

$$c'^{(RS1)}_{mn} = \lambda \frac{V'_{mn}}{E'_n - E'_m} ; \tag{11}$$

$$E^{(RS2)}_n = E'_n + \lambda \sum_{m \neq n} V'_{nm} c'^{(RS1)}_{mn} = E'_n + \lambda^2 e^{(2)}_n ;$$

$$e^{(2)}_n \equiv \sum_{m \neq n} \frac{V'_{nm} V'_{mn}}{E'_n - E'_m} . \tag{12}$$

Similarly, with the relation $E_n \cong E'_n + \lambda^2 e^{(2)}_n + O(\lambda^3)$, we can write

$$c'^{(BW2)}_{mn} = \lambda \frac{V'_{mn}}{E_n - E'_m} + \lambda^2 \sum_{m_1 \neq n} \frac{V'_{mm_1} V'_{m_1 n}}{(E_n - E'_m)(E_n - E'_{m_1})}$$

$$\cong \lambda \frac{V'_{mn}}{E'_n - E'_m} + \lambda^2 \sum_{m_1 \neq n} \frac{V'_{mm_1} V'_{m_1 n}}{(E'_n - E'_m)(E'_n - E'_{m_1})} + O(\lambda^3)$$

$$\equiv c'^{(RS2)}_{mn} + O(\lambda^3) ; \tag{13}$$

$$E^{(RS3)}_n = E'_n + \lambda \sum_{m \neq n} V'_{nm} c'^{(RS2)}_{mn} = E'_n + \lambda^2 e^{(2)}_n + \lambda^3 e^{(3)}_n ;$$

$$e^{(3)}_n \equiv \sum_{m \neq n} \sum_{m_1 \neq n} \frac{V'_{nm} V'_{mm_1} V'_{m_1 n}}{(E'_n - E'_m)(E'_n - E'_{m_1})} . \tag{14}$$

We may continue this procedure to get the higher-order approximations in RSPT. For easy reference, we will give the next order approximations:

$$c'^{(RS3)}_{mn} = c'^{(RS2)}_{mn} + \lambda^3 \left[ \sum_{m_1 \neq n} \sum_{m_2 \neq n} \frac{V'_{mm_1} V'_{m_1 m_2} V'_{m_2 n}}{(E'_n - E'_m)(E'_n - E'_{m_1})(E'_n - E'_{m_2})} - e^{(2)}_n \frac{V'_{mn}}{(E'_n - E'_m)^2} \right] ; \tag{15}$$

$$E^{(RS4)}_n = E'_n + \lambda \sum_{m \neq n} V'_{nm} c'^{(RS3)}_{mn} = E'_n + \lambda^2 e^{(2)}_n + \lambda^3 e^{(3)}_n + \lambda^4 e^{(4)}_n ;$$

$$e_n^{(4)} = \sum_{m \neq n} \sum_{m_1 \neq n} \sum_{m_2 \neq n} \frac{V'_{nm}V'_{mm_1}V'_{m_1m_2}V'_{m_2n}}{(E'_n - E'_m)(E'_n - E'_{m_1})(E'_n - E'_{m_2})} - e_n^{(2)} \sum_{m \neq n} \frac{V'_{nm}V'_{mn}}{(E'_n - E'_m)^2}. \quad (16)$$

We now propose a new iterative procedure for solving Eqs. (6) and (7) for successive approximations to the energy and the state vector:

$$c'^{(k)}_{mn} = \frac{\lambda}{E_n^{(k)} - E'_m}\left(V'_{mn} + \sum_{m_1 \neq n} V'_{mm_1} c'^{(k-1)}_{m_1 n}\right) \quad \text{for } m \neq n \quad (17)$$

$$E_n^{(k+1)} = E'_n + \lambda \sum_{m \neq n} V'_{nm} c'^{(k)}_{mn} \quad (18)$$

When $E_n$ is nondegenerate, we can take $c'^{(0)}_{mn} = 0$ for $m \neq n$, and the first-order approximation for energy is $E_n^{(1)} = E'_n$. Note again that in this paper the superscript ($k$) denotes a quantity that is correct up to $O(\lambda^k)$, not the $k$th-order correction. The first-order approximation for $c'_{mn}$ for $m \neq n$ and the second-order expression for $E_n$ are given by

$$c'^{(1)}_{mn} = \frac{\lambda V'_{mn}}{E'_n - E'_m} \quad (19)$$

$$E_n^{(2)} = E'_n + \lambda^2 \sum_{m \neq n} \frac{|V'_{nm}|^2}{E'_n - E'_m} \quad (20)$$

These expressions coincide with those of the RSPT in Eqs. (11) and (12). With Eqs. (19) and (20) it is straightforward to calculate the higher order quantities by using the recursive relations in Eqs. (17) and (18). One can see easily that in this new perturbation method the computing time required for obtaining the successive higher-order results increases linearly after the second-order calculation.

### III. QUASI-DEGENERATE PERTURBATION THEORY

The applicability of the nondegenerate perturbation theory is limited to the case where

$$|E'_m - E_n| \gg \lambda |V'_{nm}|, \tag{21}$$

for all states $m(\neq n)$. We now consider the case in which several states with the same symmetry as the $n$th state have similar energies at the level of the first-order energies; that is, $E'_1 \cong E'_2 \cong \cdots \cong E'_d$. For notational convenience, we let the $n$th state of interest be one of the $d$ states. These $d$ states must include all states for which the condition (21) is not satisfied. The secular equation to solve reads as

$$\begin{pmatrix} E'_1 - E_n & \lambda V'_{12} & \cdots & \lambda V'_{1d} & \cdots & \lambda V'_{1l} & \cdots \\ \lambda V'_{21} & E'_2 - E_n & \cdots & \lambda V'_{2d} & \cdots & \lambda V'_{2l} & \cdots \\ \vdots & \vdots & \ddots & \vdots & \vdots & \vdots & \vdots \\ \lambda V'_{d1} & \lambda V'_{d2} & \cdots & E'_d - E_n & \cdots & \lambda V'_{dl} & \cdots \\ \vdots & \vdots & \vdots & \vdots & \ddots & \vdots & \vdots \\ \lambda V'_{l1} & \lambda V'_{l2} & \cdots & \lambda V'_{ld} & \cdots & E'_l - E_n & \cdots \\ \vdots & \vdots & \vdots & \vdots & \vdots & \vdots & \ddots \end{pmatrix} \begin{pmatrix} c_{1n} \\ c_{2n} \\ \vdots \\ c_{dn} \\ \vdots \\ c_{ln} \\ \vdots \end{pmatrix} = 0 \tag{22}$$

In the case under consideration, we expect that $E'_j - E_n \sim O(\lambda^1)$ and $c_{jn} \sim O(\lambda^0)$ for $j = 1, \cdots, d$, and $E_n - E'_l \sim O(\lambda^0)$ and $c_{ln} \sim O(\lambda^1)$ for $l = d+1, d+2, \cdots$. Hence, if we neglect the terms of $O(\lambda^2)$, the equations in (22) are divided to two groups:

$$\begin{pmatrix} E'_1 - E_n & \lambda V'_{12} & \cdots & \lambda V'_{1d} \\ \lambda V'_{21} & E'_2 - E_n & \cdots & \lambda V'_{2d} \\ \vdots & \vdots & \ddots & \vdots \\ \lambda V'_{d1} & \lambda V'_{d2} & \cdots & E'_d - E_n \end{pmatrix} \begin{pmatrix} c_{1n} \\ c_{2n} \\ \vdots \\ c_{dn} \end{pmatrix} = 0, \tag{23}$$

$$\lambda \sum_{j=1}^{d} V'_{lj} c_{jn} + (E'_l - E_n) c_{ln} = 0 \quad (l = d+1, d+2, \cdots) \tag{24}$$

The secular equation (23) gives $d$ correct zeroth-order states and their first-order energies:

$$\left|\varphi_i^{(0)}\right\rangle = \sum_{j=1}^{d} c_{ji}^{(0)} \left|\psi_j^{(0)}\right\rangle \text{ and } \mathcal{E}'_i \quad (i = 1, \cdots, d), \tag{25}$$

where we will assume that $|\varphi_i^{(0)}\rangle$ is orthonormalized. The *n*th state of interest, $|\psi_n\rangle$, corresponds to the correct zeroth-order state $|\varphi_n^{(0)}\rangle$, for which the expansion coefficient $c_{nn}^{(0)}$ has the larger absolute value than other expansion coefficients.

We now take the correct zeroth-order states $\{|\varphi_j^{(0)}\rangle\}$ as the zeroth-order states instead of $\{|\psi_j^{(0)}\rangle\}$ for $j=1,\cdots,d$. To avoid confusion, we will denote the first *d* states by the index *i*, *j*, or *n*, and the later states by the index *l* or *m*, and introduce the following notations:

$$\bar{c}_{jn} = \langle\varphi_j^{(0)}|\psi_n\rangle, \quad \bar{V}'_{jm} = \langle\varphi_j^{(0)}|V|\psi_m^{(0)}\rangle, \quad \bar{V}'_{mj} = \langle\psi_m^{(0)}|V|\varphi_j^{(0)}\rangle. \tag{26}$$

Since $H_{ij} = \langle\varphi_i^{(0)}|H|\varphi_j^{(0)}\rangle = \mathcal{E}'_i\delta_{ij}$, $H_{im} = \langle\varphi_i^{(0)}|H|\psi_m^{(0)}\rangle = \lambda\bar{V}'_{im}$, and $H_{lm} = \langle\psi_l^{(0)}|H|\psi_m^{(0)}\rangle$ $= E'_m\delta_{lm} + \lambda V'_{lm}$, the secular equation (22) becomes

$$\begin{pmatrix} \mathcal{E}'_1 - E_n & 0 & \cdots & 0 & \cdots & \lambda\bar{V}'_{1l} & \cdots \\ 0 & \mathcal{E}'_2 - E_n & \cdots & 0 & \cdots & \lambda\bar{V}'_{2l} & \cdots \\ \vdots & \vdots & \ddots & \vdots & \vdots & \vdots & \vdots \\ 0 & 0 & \cdots & \mathcal{E}'_d - E_n & \cdots & \lambda\bar{V}'_{dl} & \cdots \\ \vdots & \vdots & \vdots & \vdots & \ddots & \vdots & \vdots \\ \lambda\bar{V}'_{l1} & \lambda\bar{V}'_{l2} & \cdots & \lambda\bar{V}'_{ld} & \cdots & E'_l - E_n & \cdots \\ \vdots & \vdots & \vdots & \vdots & \vdots & \vdots & \ddots \end{pmatrix} \begin{pmatrix} \bar{c}_{1n} \\ \bar{c}_{2n} \\ \vdots \\ \bar{c}_{dn} \\ \vdots \\ c_{ln} \\ \vdots \end{pmatrix} = 0 \tag{27}$$

From Eq. (27), we obtain

$$E_n = \mathcal{E}'_n + \lambda \sum_{l=d+1} \bar{V}'_{nl} c'_{ln}, \tag{28}$$

$$c'_{ln} = \frac{c_{ln}}{\bar{c}_{nn}} = \frac{\langle\psi_l^{(0)}|\psi_n\rangle}{\langle\varphi_n^{(0)}|\psi_n\rangle} = \frac{\lambda}{E_n - E'_l}\left(\bar{V}'_{ln} + \sum_{\substack{j=1 \\ j\neq n}}^{d} \bar{V}'_{lj}\bar{c}'_{jn} + \sum_{m=d+1} V'_{lm}c'_{mn}\right) \tag{29}$$

$$\bar{c}'_{jn} = \frac{\bar{c}_{jn}}{\bar{c}_{nn}} = \frac{\langle\varphi_j^{(0)}|\psi_n\rangle}{\langle\varphi_n^{(0)}|\psi_n\rangle} = \frac{\lambda}{E_n - \mathcal{E}'_j} \sum_{m=d+1} \bar{V}'_{jm}c'_{mn} \tag{30}$$

Note that these are the exact relations. In fact, the set of these three equations are equivalent to Eqs. (6) and (7), except that there is no direct mixing between the first $d$ states. From Eqs. (28) – (30), we can generate successive approximations to the energy and the state vector by iteration with $c'^{(0)}_{ln} = 0$, $\bar{c}'^{(0)}_{jn} = \bar{c}'^{(1)}_{jn} = 0$, and $E_n^{(1)} = \mathcal{E}'_n$:

$$c'^{(k)}_{ln} = \frac{\lambda}{E_n^{(k)} - E'_l} \left( \bar{V}'_{ln} + \sum_{\substack{j=1 \\ j \neq n}}^{d} \bar{V}'_{lj} \bar{c}'^{(k-1)}_{jn} + \sum_{m=d+1} V'_{lm} c'^{(k-1)}_{mn} \right) \tag{31}$$

$$\bar{c}'^{(k)}_{jn} = \frac{\lambda}{E_n^{(k)} - \mathcal{E}'_j} \sum_{m=d+1} \bar{V}'_{jm} c'^{(k-1)}_{mn} \tag{32}$$

$$E_n^{(k+1)} = \mathcal{E}'_n + \lambda \sum_{l=d+1} \bar{V}'_{nl} c'^{(k)}_{ln} \tag{33}$$

In particular, the first-order approximation for $c'_{ln}$ for $l = d+1, d+2, \cdots$ and the second-order expression for $E_n$ are given by

$$c'^{(1)}_{ln} = \frac{\lambda \bar{V}'_{ln}}{\mathcal{E}'_n - E'_l}, \tag{34}$$

$$E_n^{(2)} = \mathcal{E}'_n + \lambda^2 \sum_{l=d+1} \frac{|\bar{V}'_{nl}|^2}{\mathcal{E}'_n - E'_l} \tag{35}$$

The second-order energy expression in Eq. (35) coincide with that of the multireference perturbation theory of Chen, Davidson, and Iwata[6] [Eq. (20) of Ref. 6 truncated at the second order]. Multireference perturbation theories based on the RSPT and the BWPT with various partitioning of the Hamiltonian matrix have been widely used in molecular electronic problems.[3,5,7-10] We will show below that Eq. (35) provides quite accurate results by including just a few nearby states in constructing the correct zeroth-order states.

## IV. NUMERICAL RESULTS

We now check the accuracy of the higher-order results of the present perturbation theory against those of the BWPT and the RSPT with and without EN partitioning. The first example is a perturbed harmonic oscillator:

$$H_0 = -\frac{\hbar^2}{2\mu}\frac{d^2}{dx^2} + \frac{1}{2}\mu\omega^2 x^2, \quad \lambda V = \frac{\lambda}{2}\mu\omega^2 x^2 \tag{36}$$

Although this example must be trivial, it serves well to illustrate the quality of the approximations. The exact energy is given by $E_n^{exact} = (1+\lambda)^{1/2}\hbar\omega(n+\frac{1}{2})$. In Fig. 1(a) we display the relative errors in the ground-state energies $E_0^{(k)}$ calculated from Eq. (18). We set the values of $\hbar$, $\mu$ and $\omega$ to unity. For clarity's sake, we give only the even order results up to $E_0^{(6)}$. In fact, $E_0^{(2k+1)}$ is slightly worse than $E_0^{(2k)}$, though better than $E_0^{(2k-2)}$. Numerical results up to $k=20$ and $\lambda=50$ show that $E_0^{(k)}(\lambda)$ and $E_1^{(k)}(\lambda)$ converge to the exact values as we go to the higher order; these two states are the lowest energy states having the even and the odd parity, respectively. However, as the quantum number $n$ increases, the radius of convergence appears to decrease.

Figures 1(b) and 1(c) display the corresponding results calculated from the RSPT and the BWPT with EN partitioning, respectively. We see that both perturbation theories also provide convergent results. For this model system, the RSPT gives more accurate results than the BWPT, and the results calculated from our perturbation theory lie between those from the two theories. On the other hand, as shown in Fig. 1(d), the RSPT and the BWPT without EN partitioning give divergent results even for very small $\lambda$.

In Fig. 1(a), we also compare the results of Eq. (35) (drawn as the black dashed curve), which utilizes the correct zeroth-order states, with those obtained from Eq. (18). To calculate the ground state energy, we just mix the $|\psi_0^{(0)}\rangle$ and $|\psi_2^{(0)}\rangle$ states to obtain $|\varphi_0^{(0)}\rangle$ and $|\varphi_2^{(0)}\rangle$ states. Even with this small tweaking, the resulting second-order energy is almost as good as the straightforward fourth-order energy obtained from Eq. (18).

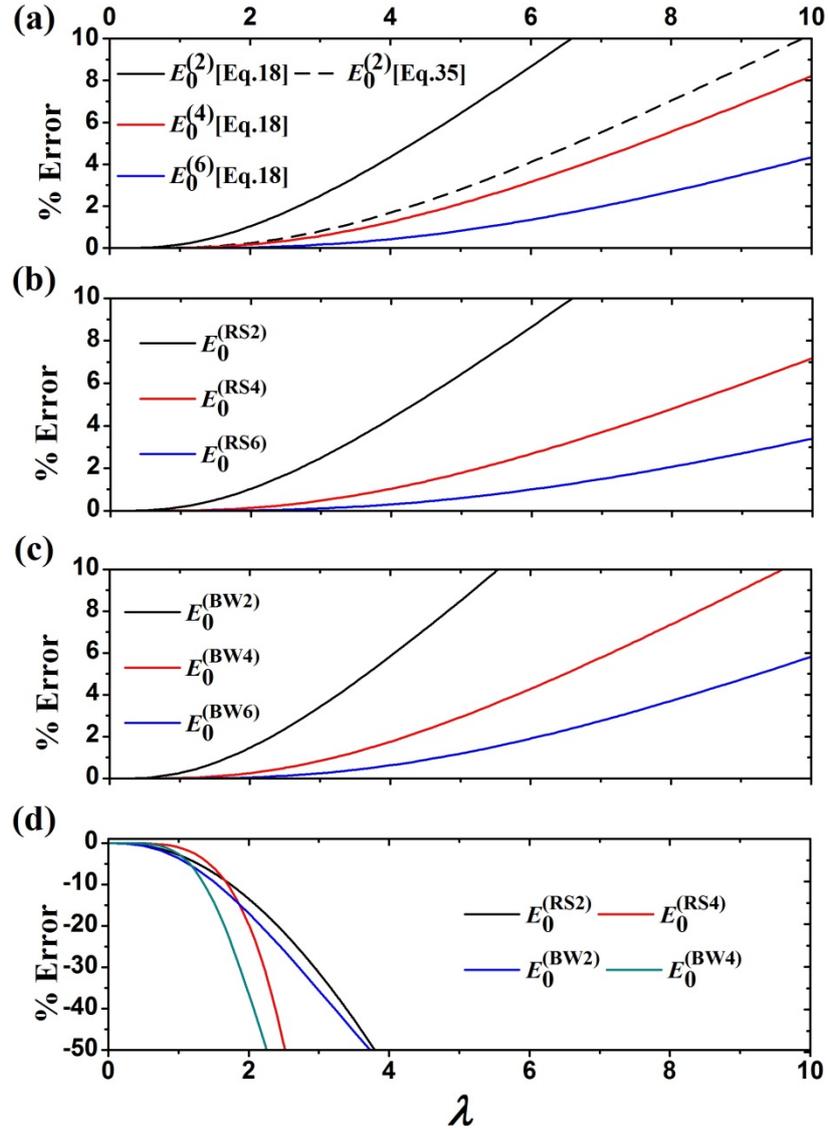

FIG. 1. Relative errors in the ground-state energies of the harmonically perturbed oscillator [Eq. (36)] calculated from the various perturbation theories as a function of the perturbation parameter $\lambda$. Figures 1(a), 1(b), and 1(c) display the results of the present, RS, and BW perturbation theories with EN partitioning, while Fig. 1(d) shows the results of the RS and BW perturbation theories without EN partitioning.

Figure 2 displays the results for a modified particle in a box with infinite walls at $x = -L/2$ and $x = L/2$. Inside the box, the potential energy is given by $V(x) = \lambda \cos(\pi x/L)$. An interesting feature of this system is that when $\lambda$ has a large positive value, $V(x)$ becomes a deep double-well potential and the ground state and the first excited state become almost degenerate, and the energy difference between the ground state and the second excited state gets small. See Fig. 2(a) for the variation of numerically calculated energy levels as a function of $\lambda$; in the calculation, the values of $\hbar$, $\mu$ and $L$ were set to unity. Hence the second-order energy calculated from Eq. (18) may become inaccurate for large positive $\lambda$ values, as shown in Fig. 2(b). In comparison, $E_0^{(2)}$ calculated from Eq. (35) is as accurate as the fourth-order energy calculated from Eq. (18) for the whole range of $\lambda$ considered; in applying Eq. (35) $|\psi_1^{(0)}\rangle$ and $|\psi_3^{(0)}\rangle$ states are mixed to obtain $|\varphi_1^{(0)}\rangle$ and $|\varphi_3^{(0)}\rangle$. Note that the downward spikes near $\lambda = 0$ result from the fact that the ground-state energy becomes zero when $\lambda$ has a small negative value.

Figures 2(c) and 2(d) display the corresponding results calculated from the RSPT and the BWPT with EN partitioning, respectively. For this model system, the results of our perturbation theory and the BWPT show better convergence behavior than those of the RSPT with EN partitioning. Again, as shown in Fig. 2(e), the RSPT and the BWPT without EN partitioning give much less accurate results; note that the ordinate scale in Fig. 2(e) is different from those of Fig. 2(b), 2(c), and 2(d).

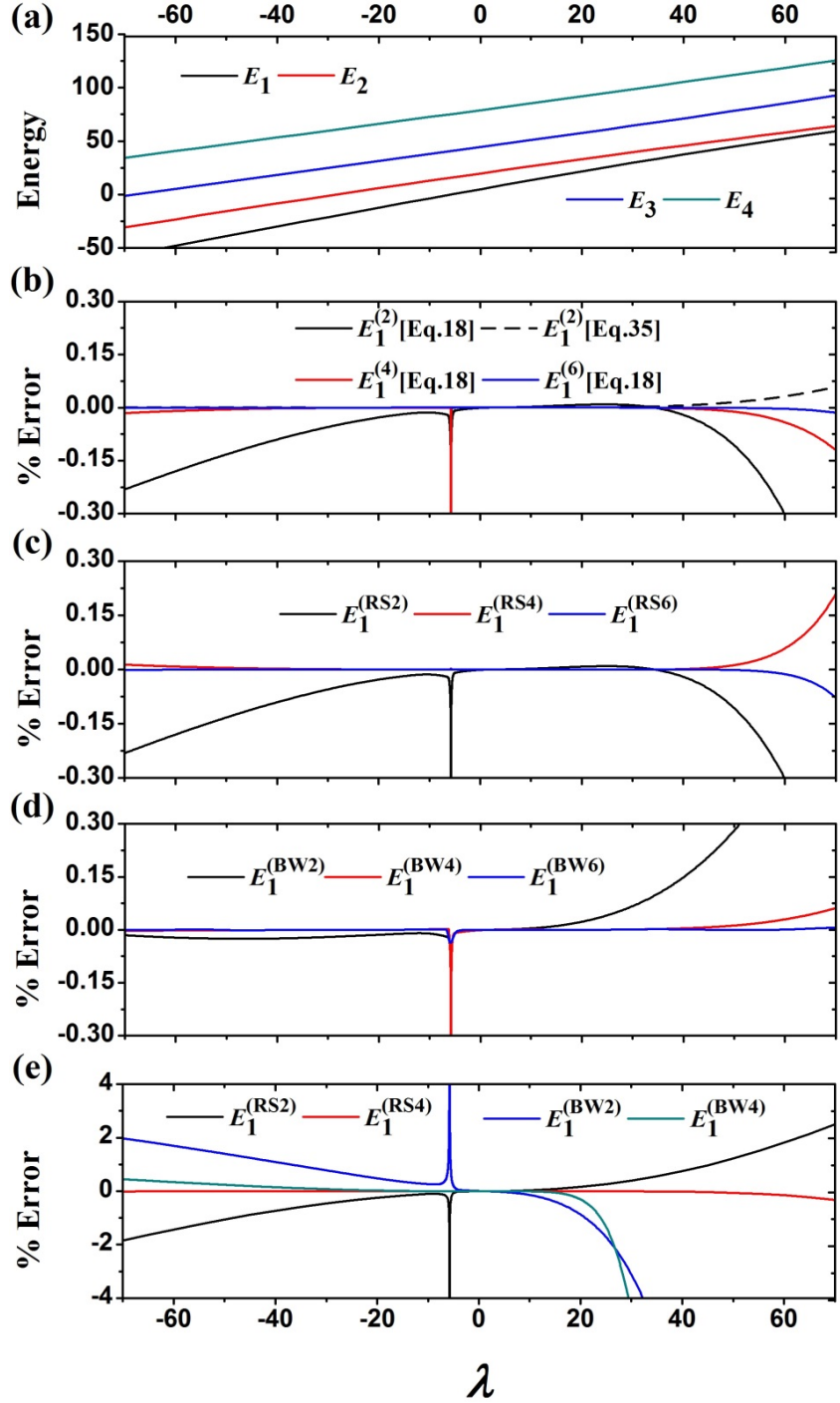

FIG. 2. (a) Variation of the numerical $E_n^{exact}(\lambda)$ values for a modified particle in a box with $V(x) = \lambda \cos(\pi x/L)$. (b) – (e) Comparison of the relative errors in the ground-state energies calculated from various perturbation theories. Figures 2(b), 2(c), and 2(d) display the results of the present, RS, and BW perturbation theories with EN partitioning, while Fig. 2(e) shows the results of the RS and BW perturbation theories without EN partitioning.

## V. CONCLUSION

We have shown that various perturbation theories with EN partitioning can be derived in a quite elementary manner. In particular, we proposed a variant of the BWPT, which can be easily extended to the quasi-degenerate case. While the usual BWPT requires rapidly increasing computing time to get the higher-order results, the new perturbation theory requires only linearly increasing computing time to get the successive higher-order results. At least for the two simple model systems considered, the new perturbation theory appears to give as accurate results as the BWPT and the RSPT with EN partitioning. The new variant BWPT would carry the similar advantages and disadvantages of the usual BWPT in comparison with the RSPT.[3] In particular, it does not satisfy the size-consistency requirement, so that it might have a limited usability in molecular electronic problems.

## ACKNOWLEDGMENTS

The authors would like to thank Prof. Haruyuki Nakano for helpful comments. This work was supported by the grants from National Research Foundation (NRF), funded by the Korean Government (Grant Nos. 2012-R1A1A2003055 and 2010-0001631).